# Two-Dimensional Semiconducting Boron Monolayers

Shao-Gang Xu,[1,2,#] Xiao-Tian Li,[1,#] Yu-Jun Zhao,[1] Ji-Hai Liao,[1] Xiao-Bao Yang,[1,2,*] and Hu Xu[2,*]

[1]*Department of Physics, South China University of Technology, Guangzhou 510640, People's Republic of China*

[2]*Department of Physics, South University of Science and Technology of China, Shenzhen 518055, People's Republic of China*

The two-dimensional boron monolayers were reported to be metallic both in previous theoretical predictions and experimental observations, however, we have firstly found a family of boron monolayers with the novel semiconducting property as confirmed by the first-principles calculations with the quasi-particle $G_0W_0$ approach. We demonstrate that the vanished metallicity characterized by the $p_z$-derived bands cross the Fermi level is attributed to the motif of a triple-hexagonal-vacancy, with which various semiconducting boron monolayers are designed to realize the band-gap engineering for the potential applications in electronic devices. The semiconducting boron monolayers in our predictions are expected to be synthesized on the proper substrates, due to the similar stabilities to the ones observed experimentally.

Graphene[1], a free-standing two-dimensional (2D) Carbon (C) exfoliated from the bulk phase, has provoked a rush of attentions. However, the intrinsic metallicity of graphene prevents its application for practical nano-devices, and various means have been proposed to introduce a suitable band gap for graphene, such as the chemical functionalization (hydrogenated graphene)[2] and physical cutting (graphene nanoribbons)[3,4]. Recently, the few-layer black phosphorus is considered as a better choice to achieve field-effect transistor (FET) performances at room temperature, due to their semiconducting characteristic[5]. Most reported 2D materials are attractive due to their novel properties, however, the atomic structures can be theoretically determined through the relaxation from the layered bulk counterpart[6].

Boron(B), one of C's nearest neighbors in the periodic table, demonstrates the polymorphism in low-dimensional structures[7-9] compared to the bulk, due to the multi-centre bonds. Experimental observations show that small $B_n$ clusters are planar triangular fragments[10,11] which satisfy the Aufbau principle[12], however, the corresponding B monolayers have been found to be not stable due to the large deformation and buckling[13]. Inspired by the computational prediction of $B_{80}$[14], the stable B monolayers with hexagonal vacancies have been proposed[15,16], leading to extensive searches focusing on the concentration and distribution of vacancies[17-20]. Further experimental observation of $B_{35}^-$[21] and $B_{36}$[22] confirm that hexagonal vacancies are important to stabilize the larger planar B clusters as well as the two-dimensional borophene[22,23], which are supposed to be synthesized on the proper metal surfaces[24,25]. Recently, two research groups have independently reported that the borophene could be fabricated on the silver substrates by the molecular beam epitaxy (MBE) method[26,27], with the detailed atomic structures confirmed by the first-principles calculations[28-30].

As is known, the polymorphism of boron monolayers will induce the variety of properties, including the unusual Dirac fermions[31], and the coexisting Dirac nodal lines and tilted semi-Dirac cones[32]. During the growth of borophene, the metallic B nanoribbons (BNRs) have been found across the steps on the silver substrates[33,34]. To our knowledge, there are only few B periodical structures predicted to be semiconducting. In addition to the one-dimensional (1D) B nanotubes with small diameters[16], only one intrinsic BNRs with armchair edges[35,36] has been predicted to be semiconducting, and the 1D B chains[37] will undergo a metal-semiconductor transition due to the mechanism of spin density wave. Till now, the planar B monolayers are concluded to be metallic[9,17,38] by experiments and theories, characterized by the out-of-plane $p_z$-derived bands. However, it is not clear whether there are any semiconducting boron monolayers, the confirmation of which may be an important complementary to the previous literature.

In this work, we have theoretically shown that the $p_z$-derived bands in boron monolayers could be effectively modulated by a triple-hexagonal-vacancy (THV), which results in a family of semiconducting B monolayers with indirect and direct gaps as confirmed by the first-principles calculations and the quasi-particle $G_0W_0$ approach. We have found a simple rule to construct semiconducting B monolayers by the combination of BNRs with the THV chains for the realization of band-gap engineering, indicating a brand new platform for the potential applications in electronic devices.

The first-principles calculations were performed based on density-functional theory (DFT) with the projected augmented wave (PAW)[39] scheme, as implemented in Vienna *ab initio* simulation package (VASP)[40,41]. The generalized gradient approximations (GGA)[42,43] expressed by the Perdew-Burke-Ernzerhof (PBE) functional were adopted, with a 480 eV energy cutoff for the plane-wave basis set. Both the lattice constants and atomic positions were fully optimized with the convergence of force on each atom less than 0.01 eV/Å. To avoid the interaction between adjacent images, a vacuum region of 21 Å along the $z$ direction was added for the model. In addition, the hybrid functional[44] HSE06 was also employed to confirm the structural stabilities and the electronic properties of the potential semiconducting structures. For the monolayers with direct gaps, the quasi-particle $G_0W_0$ approach[45] was applied for the band gap corrections. We also used the DMol$^3$[46,47] method combined with the exchange-correlation functional of GGA (PBE) to study the electronic properties of the B clusters. To confirm the dynamical stability, the phonon band structures were calculated with the finite displacement method as implemented in the Phonopy program[48], where the precision convergence criteria for the total energy is $10^{-8}$ eV. Thermal stability was also studied using *ab initio* molecular dynamics (AIMD) simulations for the supercell with the temperature controlled by a Nosé heat bath scheme[49].

In our recent study, we have found a highly stable $B_{49}^+$ cluster with an energy gap of 1.2 eV[50], where a double-hexagonal-vacancy (DHV) located at the center (shown in Fig. S1) results in the energy gap opening of the out-of-plane $p_z$ orbitals. Inspired by this, we have constructed a highly symmetrical ($C_{3v}$) $B_{33}^-$ cluster with a triple-hexagonal-vacancy (THV), and the projected density of states (PDOS) show that the lowest unoccupied molecular orbitals (LUMO) level is attributed to the in-plane (sum of $s$, $p_x$ and $p_y$) $s+p_{x,y}$ orbitals rather than the out-of-plane $p_z$ orbitals which dominate the highest occupied molecular orbitals (HOMO) level (shown in Fig. 1(a)). Note that all the stable B monolayers previously proposed are metallic[15-20] as characterized by the $p_z$-derived bands, where the Fermi level ($E_F$) falls in the gap of the in-plane states ($s+p_{x,y}$). Thus, the introducing of THVs might be helpful to achieve the semiconducting monolayers by opening the energy gap of $p_z$ orbitals.

As shown in Fig. 1(b), we have constructed a novel B monolayer ($\beta_0^s$) with the THV units and triangular regions, where the real space structural parameters after relaxation are shown in Table S1. The electronic band structure calculated by the HSE06 functional (shown in Fig. 1(c)) confirms that this monolayer is semiconducting with an indirect gap of 0.741 eV. The projected band structure reveals that, the valence band maximum (VBM) is dominated by the out-of-plane $p_z$ orbitals while the conduction band minimum (CBM) is attributed to the in-plane $s+p_{x,y}$ orbitals, which are similar with the HOMO-LUMO analysis of $B_{33}^-$ cluster. According to the calculated phonon dispersion (shown in Fig. 1(d)), no imaginary phonon modes are found in the whole first Brillouin zone, indicating the kinetic stability of the new B monolayer. We also assessed the thermal stability of the B monolayer by performing AIMD simulation at the temperature of 500 K with the time step is 1 fs. The structural snapshot of the supercell at 10 ps (shown in Fig. 1(e)) reveals that the semiconducting monolayer has a good thermodynamic stability, which can maintain the structural integrity in room-temperature environment.

To explore other possible semiconducting B monolayers, we have screened candidates based on the triangle lattice with hexagon vacancies, where the unit cells contain THV motifs with the number of B atoms less than 30. Considering the structural stabilities, the vacancy concentration in the monolayers is limit from 1/9 to 1/5, and all B atoms' coordination numbers are not smaller than 4. The semiconducting B monolayers can be decomposed into the triangle regions, hexagonal regions and the parallel THV chains, where the hexagonal vacancies are necessary to maintain the semiconducting characteristic as the ratio of the triangle regions increases. According to the structures of the three semiconducting B monolayers with indirect-gaps shown in Fig. S2, we can conclude that the THV chains are critical for the gap opening in the B monolayers.

Among the semiconducting monolayers found in our screening, there is a typical class with the motifs of the THV chains and the nanoribbons from β sheet[15], indicating a possible way to construct a family potential semiconducting B monolayers by the variation of BNRs' width (shown in Fig. 2(a)). The named $\beta_m^s$ monolayer is one with widths($m$) of the BNRs, where there are no BNRs in the unit cell of the $\beta_0^s$ monolayer (shown in Fig. 1(b)) with the least number of atoms. Under the above definition, we can form the $\beta_1^s$ monolayer with the BNRs (*m*=1) assembled between the THV chains, as shown in Fig. 2b, and the $\beta_2^s$ monolayer in the Fig. 2(c) is created in the same way. Furthermore, we attempt to assemble the BNRs of various widths with the THV chains, where the $\beta_{1,2}^s$ monolayer (shown in Fig. 2(d)) is the hybrid structures of $\beta_1^s$ and $\beta_2^s$. The structural parameters of these three new monolayers with the same space group of *Pm*(6) are presented in Table S1. The results of phonon dispersions along the high-symmetry lines and the AIMD simulations at the temperature of 500 K for the corresponding structures are shown in Fig. S3, which indicate that the three monolayers should have high dynamic and thermal stabilities.

The HSE06 band structures (shown in Fig. 3(a-c)) show that the above B monolayers are semiconducting with the band gaps of 0.602, 0.416 and 0.396 eV, respectively. The $\beta_1^s$ monolayer is an indirect gap semiconductor with the VBM is along M-Γ, and the CBM is located at Γ point. Notably, the $\beta_2^s$ monolayer is a semiconductor with direct gap, where both the VBM and CBM is located at Y point. The $\beta_{1,2}^s$ monolayer is an indirect gap semiconductor with the VBM is along Γ-X, and the CBM is located at Y point. Checking the projected orbitals plotted in the band structures, we can see that, the VBM and CBM are contributed by the in-plane $s+p_{x,y}$ orbitals for the $\beta_1^s$, $\beta_2^s$ and $\beta_{1,2}^s$ monolayers. Different from the previous B sheets where the $p_z$ orbitals account for the metallic properties, and the above B monolayers in our prediction open the energy gaps of $p_z$ orbitals near the $E_F$ level.

As the width of BNRs in the unit cell increases, the band structures and the PDOS analysis (shown in Fig. S4) indicates that the systems become metallic gradually, where the bands near the $E_F$ level are attributed to the $p_z$ orbitals rather than the in-plane $s+p_{x,y}$ orbitals. According to the charge distributions for the bands near the $E_F$ level (shown in Fig. S5), the charge of the valence bands are moving from the THV regions to the β BNRs regions as the increasing of the BNRs' width, which leads to the closing of the energy gap for $p_z$ orbitals in $\beta_m^s$ monolayers with $m \geqslant 5$. By the combinations ($\beta_{m1,m2,m3...}^s$ ($m_i$=0~4)) in the unit cell, we have construct a family of semiconducting B monolayers with various band gaps (shown in Table S2), indicating the realization of the band-gap engineering. As shown in Fig. 3(d), the $E_{\text{tot}}$ increased linearly as a function of the vacancy concentration (η), and at a fixed η we can construct different structures with various band gaps. In addition, the average total energy ($E_{\text{tot}}$) of a few semiconducting B monolayers are in the region between the $β_{12}$ (-6.229 eV/atom), $χ_3$ (-6.245 eV/atom) sheets[27] available experimentally, indicating the possible synthesis on the proper substrates.

Similar to the few-layer black phosphorus which bridges the gap between graphene and transition metal dichalcogenides[51], the direct band gap characteristic of $\beta_2^s$ monolayer can be potentially applied in the infrared optoelectronic devices and FET. The $G_0W_0$ calculation (shown in Fig. 4(a)) predicts that the fundamental band gap of $\beta_2^s$ monolayer is 0.484 eV, which maintains the characteristic of direct gap at Y point with a linear increasing as the biaxial strain ($-3\% \leqslant σ \leqslant 3\%$). The semiconducting characteristic in B monolayers is expected to be robust against the strain induced by the substrate and the band gaps could be effectively modulated by means of the biaxial strain. As shown in Fig. 4(b), the band decomposed charge distributions of VBM and CBM for $\beta_2^s$ are found to be mainly contributed to the in-plane $sp^2$ hybrid orbitals, which are located in the region of THV chains. Therefore, we confirm that the vanished metallicity characterized by the $p_z$-derived bands cross the Fermi level is attributed to the motif of THV chains in the semiconducting B monolayers.

In summary, we have firstly demonstrated a family of boron monolayers with the novel semiconducting property, attributed to the $p_z$-derived bands modulated by the triple-hexagonal-vacancy. A simple rule is proposed to design a special group of semiconducting B monolayers with various gaps, indicating a feasible band-gap engineering in a relative big region (0.2~0.9 eV) which may act as a bridge between graphene and phosphorene. The semiconducting B monolayers in our prediction may be synthesized on the proper substrates, since they are of similar stability to those observed experimentally. For the monolayer with direct gap, mechanical biaxial strains within a small region may induce a linear tuning of the band gaps, indicating a brand new platform for the potential applications in electronic devices.

**ACKNOWLEDGMENTS**

This work was supported by the National Natural Science Foundation of China (Nos. 11474100, 11674148, 11574088), Guangdong Natural Science Funds for Distinguished Young Scholars (No.2014A030306024), and the Basic Research Program of Science, Technology and Innovation Commission of Shenzhen Municipality (Grant No. JCYJ20160531190054083).

**Notes:**
[#]These authors contributed equally to this work. The authors declare no competing financial interest.
*scxbyang@scut.edu.cn; xu.h@sustc.edu.cn

**REFERENCE:**

[1]  A. H. Castro Neto, F. Guinea, N. M. R. Peres, K. S. Novoselov, and A. K. Geim, Rev. Mod. Phys. **81**, 109 (2009).
[2]  G. Fiori, S. Lebègue, A. Betti, P. Michetti, M. Klintenberg, O. Eriksson, and G. Iannaccone, Phys. Rev. B **82**, 153404 (2010).
[3]  Y.-W. Son, M. L. Cohen, and S. G. Louie, Phys. Rev. Lett. **97**, 216803 (2006).
[4]  C. Stampfer, J. Güttinger, S. Hellmüller, F. Molitor, K. Ensslin, and T. Ihn, Phys. Rev. Lett. **102**, 056403 (2009).
[5]  L. K. Li, Y. J. Yu, G. J. Ye, Q. Q. Ge, X. D. Ou, H. Wu, D. L. Feng, X. H. Chen, and Y. B. Zhang, Nat. Nanotech. **9**, 372 (2014).
[6]  G. R. Bhimanapati *et al.*, ACS Nano **9**, 11509 (2015).
[7]  L. S. Wang, Int. Rev. Phys. Chem. **35**, 69 (2016).

[8] A. I. Boldyrev and L. S. Wang, Phys. Chem. Chem. Phys. **18**, 11589 (2016).

[9] Z. Zhang, E. S. Penev, and B. I. Yakobson, Nat Chem **8**, 525 (2016).

[10] H. J. Zhai, B. Kiran, J. Li, and L. S. Wang, Nat Mater **2**, 827 (2003).

[11] A. N. Alexandrova, A. I. Boldyrev, H.-J. Zhai, and L.-S. Wang, Coord. Chem. Rev. **250**, 2811 (2006).

[12] I. Boustani, Phys. Rev. B **55**, 16426 (1997).

[13] J. Kunstmann and A. Quandt, Phys. Rev. B **74**, 035413 (2006).

[14] N. Gonzalez Szwacki, A. Sadrzadeh, and B. I. Yakobson, Phys. Rev. Lett. **98**, 166804 (2007).

[15] H. Tang and S. Ismail-Beigi, Phys. Rev. Lett. **99**, 115501 (2007).

[16] X. Yang, Y. Ding, and J. Ni, Phys. Rev. B **77**, 041402 (2008).

[17] E. S. Penev, S. Bhowmick, A. Sadrzadeh, and B. I. Yakobson, Nano Lett. **12**, 2441 (2012).

[18] X. Wu, J. Dai, Y. Zhao, Z. Zhuo, J. Yang, and X. C. Zeng, ACS Nano **6**, 7443 (2012).

[19] X. Yu, L. Li, X.-W. Xu, and C.-C. Tang, J. Phys. Chem. C **116**, 20075 (2012).

[20] H. Lu, Y. Mu, H. Bai, Q. Chen, and S. D. Li, J. Chem. Phys. **138**, 024701 (2013).

[21] W. L. Li *et al.*, J. Am. Chem. Soc. **136**, 12257 (2014).

[22] Z. A. Piazza, H. S. Hu, W. L. Li, Y. F. Zhao, J. Li, and L. S. Wang, Nat. Commun. **5**, 3113 (2014).

[23] A. P. Sergeeva, I. A. Popov, Z. A. Piazza, W. L. Li, C. Romanescu, L. S. Wang, and A. I. Boldyrev, Acc. Chem. Res. **47**, 1349 (2014).

[24] Y. Liu, E. S. Penev, and B. I. Yakobson, Angew. Chem. Int. Ed. Engl. **52**, 3156 (2013).

[25] Z. Zhang, Y. Yang, G. Gao, and B. I. Yakobson, Angew. Chem. Int. Ed. Engl. **53**, 13022 (2015).

[26] A. J. Mannix *et al.*, Science **350**, 1513 (2015).

[27] B. J. Feng *et al.*, Nat. Chem. **8**, 564 (2016).

[28] S. G. Xu, Y. J. Zhao, J. H. Liao, X. B. Yang, and H. Xu, Nano Res. **9**, 2616 (2016).

[29] H. Shu, F. Li, P. Liang, and X. Chen, Nanoscale **8**, 16284 (2016).

[30] T. Tsafack and B. I. Yakobson, Phys. Rev. B **93**, 165434 (2016).

[31] B. Feng *et al.*, Phys. Rev. Lett. **118**, 096401 (2017).

[32] H. Zhang, Y. Xie, Z. Zhang, C. Zhong, Y. Li, Z. Chen, and Y. Chen, J. Phys. Chem. Lett. **8**, 1707 (2017).

[33] Z. Zhang, A. J. Mannix, Z. Hu, B. Kiraly, N. P. Guisinger, M. C. Hersam, and B. I. Yakobson, Nano Lett. **16**, 6622 (2016).

[34] Q. Zhong *et al.*, Phys. Rev. Materials **1**, 021001 (2017).

[35] Y. Ding, X. Yang, and J. Ni, Appl. Phys. Lett. **93**, 043107 (2008).

[36] S. Saxena and T. A. Tyson, Phys. Rev. Lett. **104**, 245502 (2010).

[37] M. Liu, V. I. Artyukhov, and B. I. Yakobson, J. Am. Chem. Soc. **139**, 2111 (2017).

[38] E. S. Penev, A. Kutana, and B. I. Yakobson, Nano Lett. **16**, 2522 (2016).

[39] G. Kresse and D. Joubert, Phys. Rev. B **59**, 1758 (1999).

[40] G. Kresse and J. Hafner, Phys. Rev. B **47**, 558 (1993).

[41] G. Kresse and J. Furthmüller, Phys. Rev. B **54**, 11169 (1996).

[42] J. P. Perdew, K. Burke, and M. Ernzerhof, Phys. Rev. Lett. **77**, 3865 (1996).

[43] J. P. Perdew, K. Burke, and M. Ernzerhof, Phys. Rev. Lett. **78**, 1396 (1997).

[44] J. Heyd, G. E. Scuseria, and M. Ernzerhof, J. Chem. Phys. **118**, 8207 (2003).

[45] M. Shishkin and G. Kresse, Phys. Rev. B **74**, 035101 (2006).

[46] B. Delley, J. Chem. Phys. **92**, 508 (1990).

[47] B. Delley, J. Chem. Phys. **113**, 7756 (2000).

[48] A. Togo, F. Oba, and I. Tanaka, Phys. Rev. B **78**, 134106 (2008).
[49] S. Nosé, J. Chem. Phys. **81**, 511 (1984).
[50] S.-G. Xu, Y.-J. Zhao, J.-H. Liao, and X.-B. Yang, J. Chem. Phys. **142**, 214307 (2015).
[51] A. Castellanos-Gomez, J. Phys. Chem. Lett. **6**, 4280 (2015).

**Figure Captions**

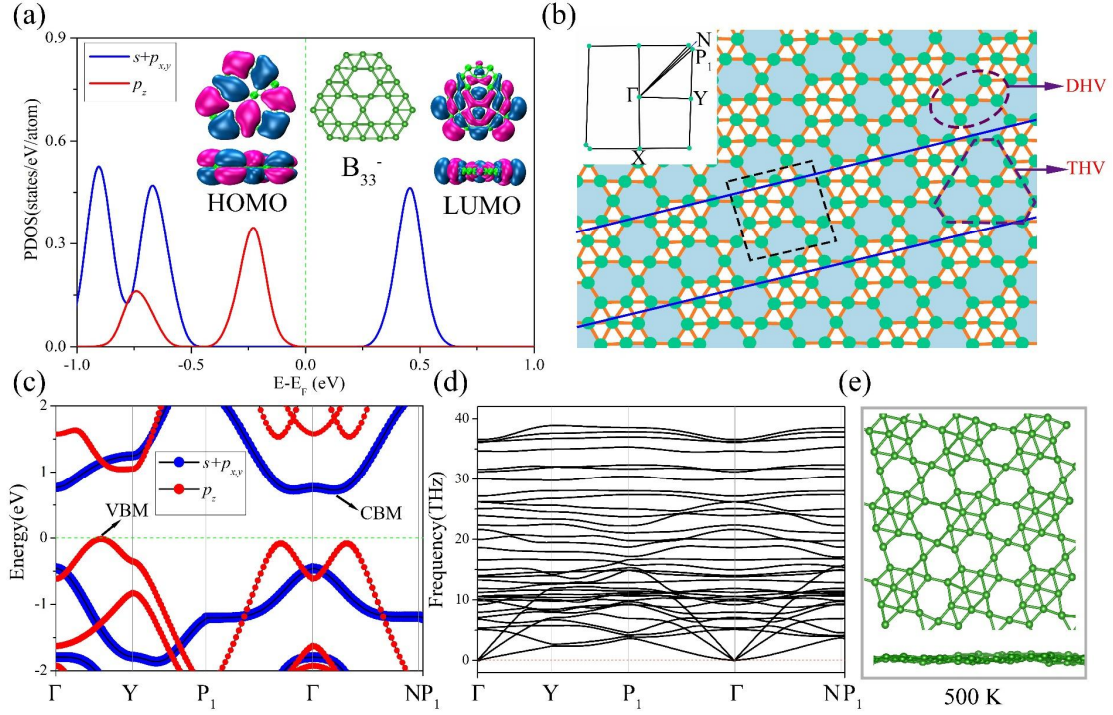

FIG. 1 (color online). The structures and electronic properties of $B_{33}^-$ cluster and the $\beta_0^s$ monolayer. (a) The PDOS analysis for the $B_{33}^-$ cluster, and the insets represent the structure and HOMO (LUMO) charge distributions of the $B_{33}^-$ cluster. (b) and (c) represent the atomic structure and HSE06 band structure of $\beta_0^s$ monolayer, respectively. (VBM is set to 0.) (d) The phonon dispersion of $\beta_0^s$ monolayer. (e) The AIMD snapshot of the $\beta_0^s$ monolayer at 10 ps of 500 K.

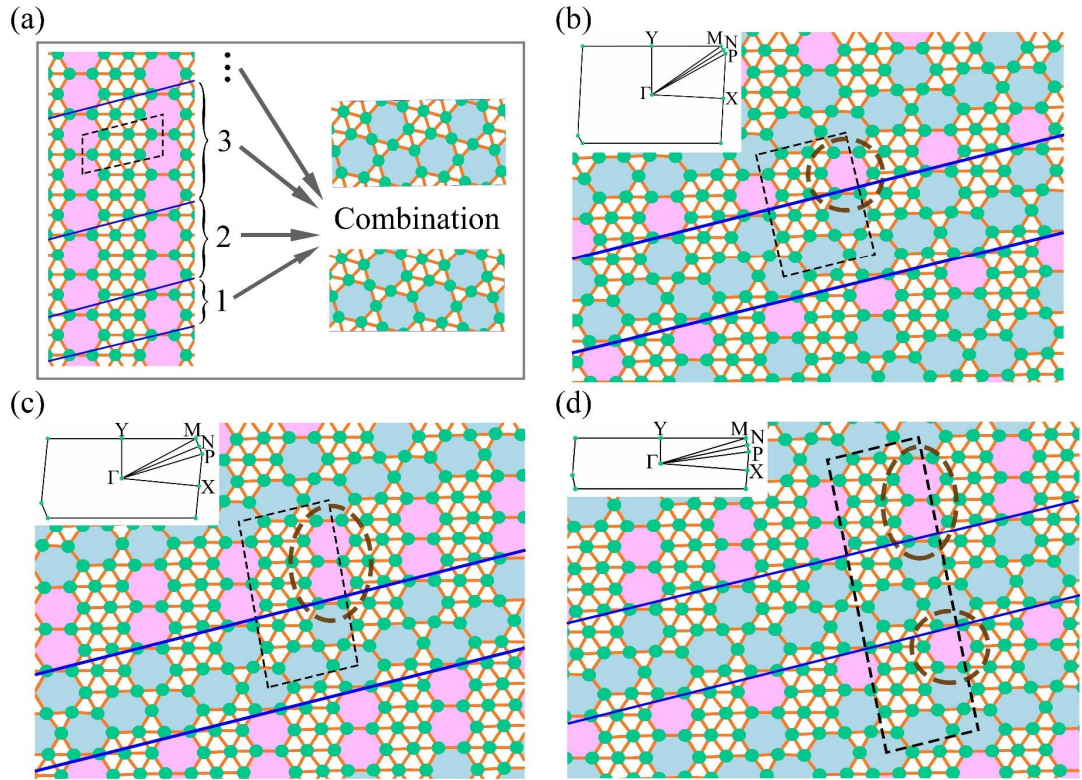

FIG. 2 (color online). The combination rule and atomic structures of the semiconducting B monolayers. (a) Schematic diagram of the combination rule for the semiconducting B monolayers. (b-d) represent the geometry structures and the first Brillouin-zones (up insets) of $\beta_1^s$, $\beta_2^s$ and $\beta_{1,2}^s$ monolayers, respectively.

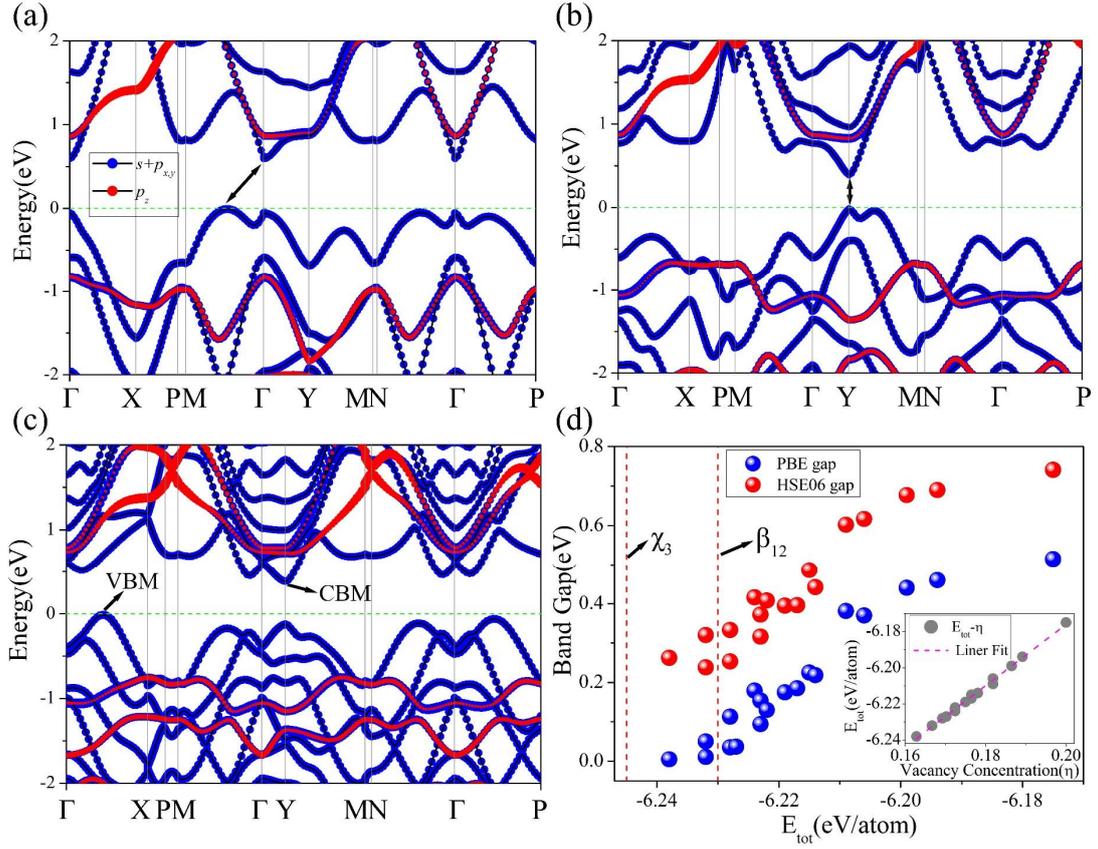

FIG. 3 (color online). The band structures and relation of band gaps and average total energy for the semiconducting B monolayers. (a-c) The HSE06 band structures for $\beta_1^s$, $\beta_2^s$ and $\beta_{1,2}^s$ monolayers, respectively. (VBM is set to 0). (d) The relationship of band gaps and average total energy ($E_{tot}$) for the new designed semiconducting B monolayers, the inset represents the relationship of $E_{tot}$ and vacancy concentration (η).

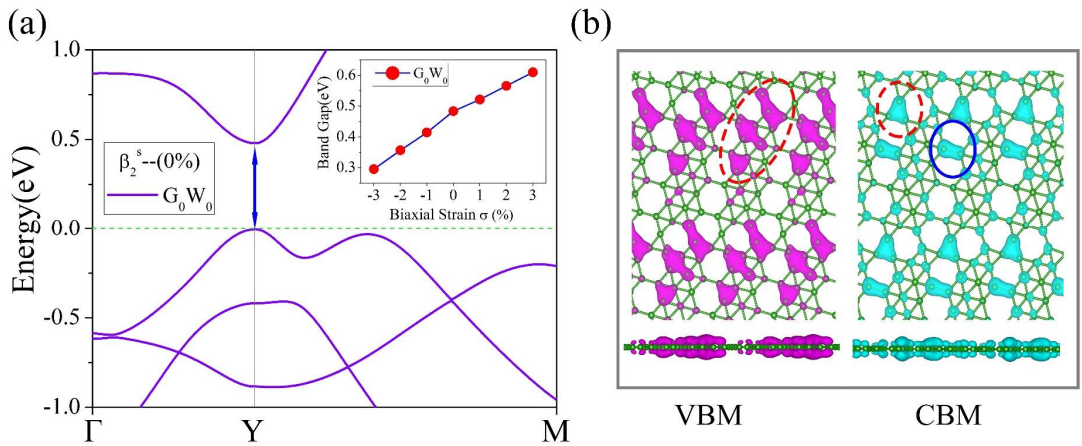

FIG. 4 (color online). The relation of band gaps and biaxial strain for $\beta_2^s$ monolayer. (a) The $G_0W_0$ band structure of $\beta_2^s$ monolayer (VBM is set to 0), and the inset represents the relation of $G_0W_0$ band gaps and biaxial strain. (b) The band decomposed charge densities of $\beta_2^s$ monolayer at VBM and CBM.

# Supplemental Material

# Two-Dimensional Semiconducting Boron Monolayer

Shao-Gang Xu,[1,2,#] Xiao-Tian Li,[1,#] Yu-Jun Zhao,[1] Ji-Hai Liao,[1] Xiao-Bao Yang,[1,2,*] and Hu Xu[2,*]

[1]*Department of Physics, South China University of Technology, Guangzhou 510640, People's Republic of China*

[2]Department of Physics, South University of Science and Technology of China, Shenzhen 518055, People's Republic of China

*E-mail: scxbyang@scut.edu.cn; xu.h@sustc.edu.cn*

Table S1. Symmetry, optimized lattice parameters, hexagonal vacancy concentration (η), and the average total energy ($E_{tot}$) of the B monolayers form GGA(PBE) results.

| Phase | Space Group | $a$ (Å) | $b$ (Å) | $\gamma$ (°) | η | $E_{tot}$ (eV/atom) |
|---|---|---|---|---|---|---|
| $\beta_0^s$ | $Amm2$(38) | 6.111 | 6.111 | 91.87 | 1/5 | -6.175 |
| $\beta_1^s$ | $Pm$(6) | 6.057 | 8.889 | 86.99 | 2/11 | -6.209 |
| $\beta_2^s$ | $Pm$(6) | 6.057 | 11.760 | 84.32 | 5/29 | -6.224 |
| $\beta_{1,2}^s$ | $Pm$(6) | 6.064 | 20.669 | 85.47 | 9/51 | -6.217 |

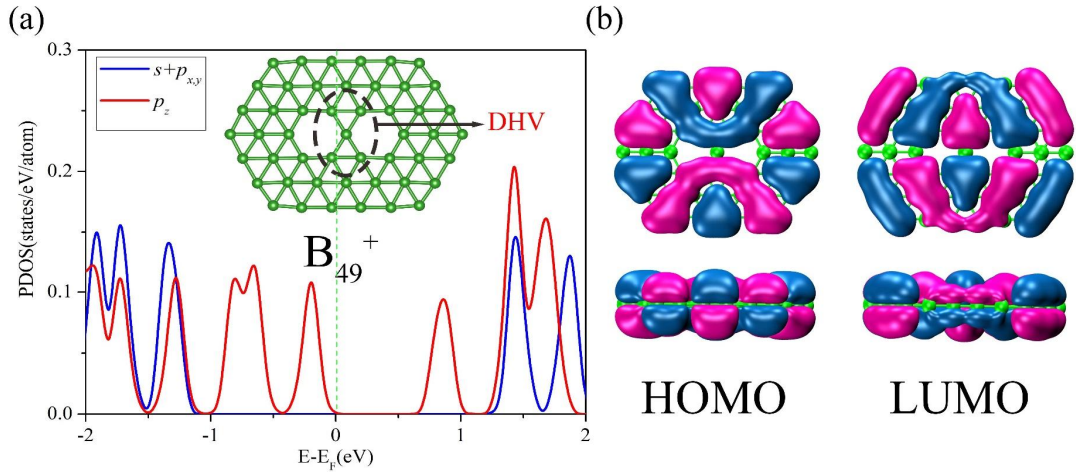

FIG. S1. (a) The PDOS analysis (HSE06) and (b) the HOMO and LUMO charge distributions for $B_{49}^+$ cluster.

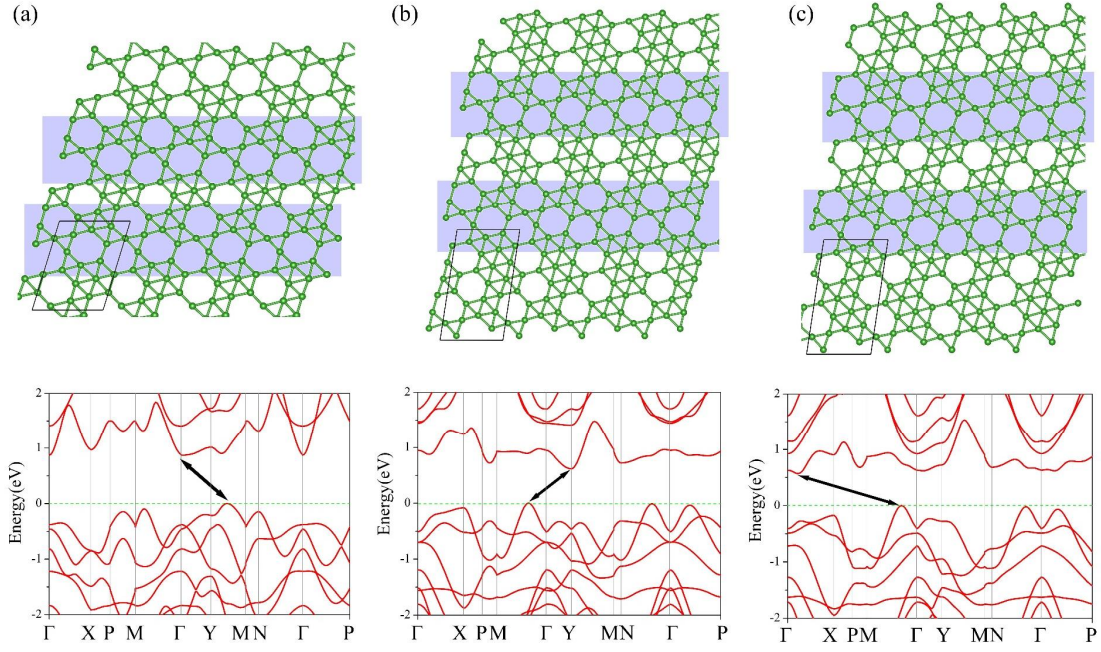

FIG. S2. The atomic structures and HSE06 band structures for new B monolayer semiconductors. (a-c) represent the results of $\beta^s_{a1}$, $\beta^s_{a2}$ and $\beta^s_{a3}$ monolayers, respectively. The blue regions represent the THV chains. The black arrows represent the position of the VBM and CBM of the corresponding monolayers. (The VBM is set to 0.)

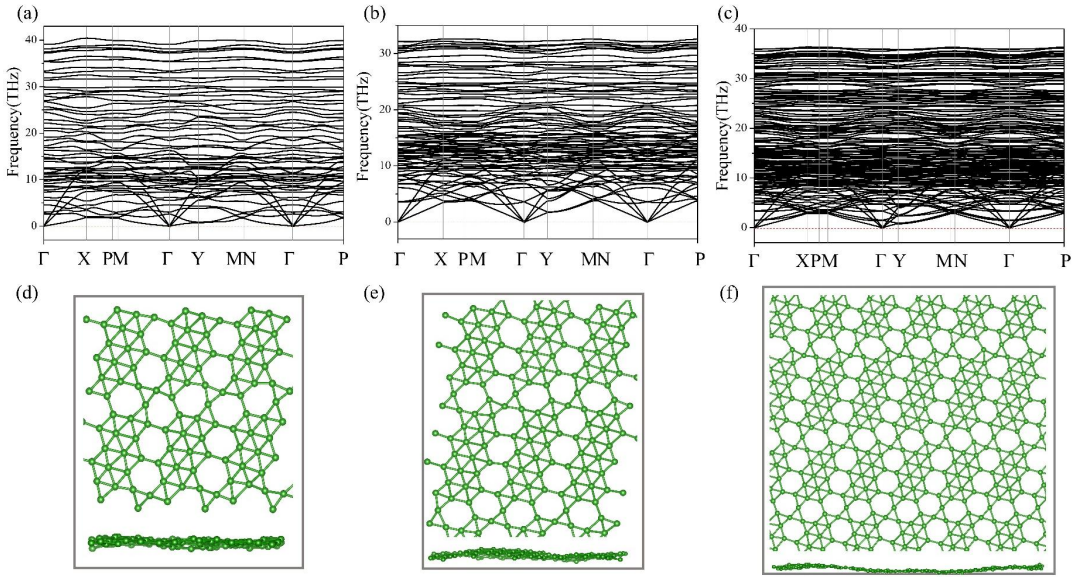

FIG. S3. (a-c) represent the phonon dispersions along the high-symmetry line of the $\beta^s_1$, $\beta^s_2$ and $\beta^s_{1,2}$ monolayers, respectively. (d-f) represent the AIMD snapshots at the temperature of 500 K (10 ps) for the $\beta^s_1$, $\beta^s_2$ and $\beta^s_{1,2}$ monolayers, respectively.

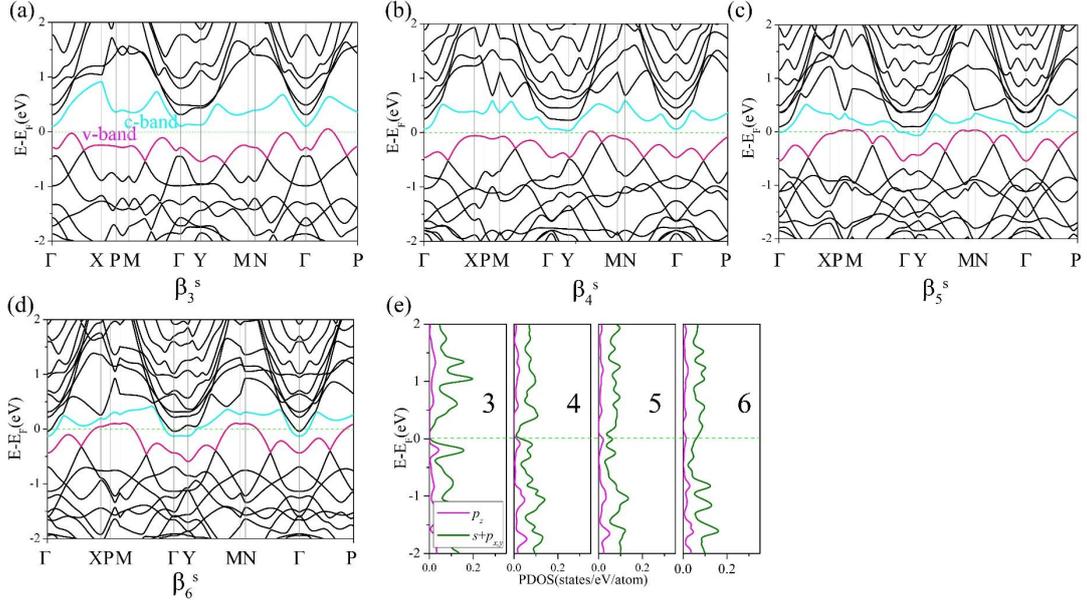

FIG. S4. (a-d) The band structures for the four B monolayers ($\beta^s_m$, $m$=3~6) at PBE level. (e) The PDOS analysis results of the above monolayers.

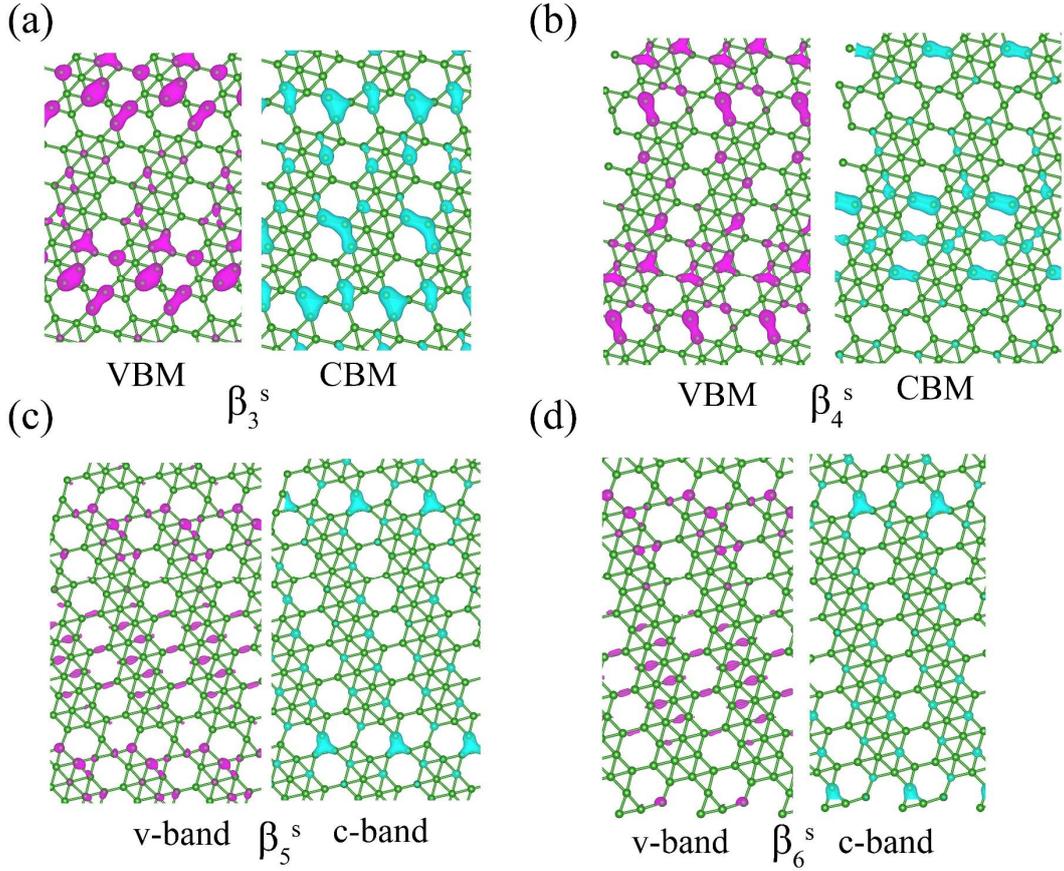

FIG. S5. (a,b) The band decomposed charge densities of the VBM and CBM for $\beta^s_3$, $\beta^s_4$ monolayers. (c,d) represent the charge densities for the valance band (v-band) and conduction band (c-band) of $\beta^s_5$, $\beta^s_6$ monolayers, respectively.

Table S2. The number of B atoms for various coordination numbers (CN=4-6), the total number (Num) of the atoms in the unit cell, the band gaps calculated by PBE and HSE06 functional, the vacancy concentration ($\eta$) and the average total energy ($E_{tot}$) of the new B monolayers combined by THV chains and $\beta$ BNRs are listed.

| Phase | (CN=4) | (CN=5) | (CN=6) | Num | Bandgap-PBE[HSE06](eV) | | $\eta$ | $E_{tot}$(eV/atom)(PBE) |
|---|---|---|---|---|---|---|---|---|
| $\beta_0^s$ | 7 | 4 | 1 | 12 | 0.515 | 0.741 | 1/5 | -6.175 |
| $\beta_1^s$ | 9 | 6 | 3 | 18 | 0.381 | 0.602 | 2/11 | -6.209 |
| $\beta_2^s$ | 11 | 8 | 5 | 24 | 0.179 | 0.416 | 5/29 | -6.224 |
| $\beta_3^s$ | 13 | 10 | 7 | 30 | 0.05 | 0.320 | 1/6 | -6.232 |
| $\beta_4^s$ | 15 | 12 | 9 | 36 | 0.005 | 0.262 | 7/43 | -6.238 |
| $\beta_{0,1}^s$ | 16 | 10 | 4 | 30 | 0.460 | 0.690 | 7/37 | -6.194 |
| $\beta_{0,2}^s$ | 18 | 12 | 6 | 36 | 0.370 | 0.617 | 2/11 | -6.206 |
| $\beta_{1,2}^s$ | 20 | 14 | 8 | 42 | 0.185 | 0.396 | 9/51 | -6.217 |
| $\beta_{0,3}^s$ | 20 | 14 | 8 | 42 | 0.225 | 0.486 | 9/51 | -6.215 |
| $\beta_{0,4}^s$ | 22 | 16 | 10 | 48 | 0.130 | 0.408 | 5/29 | -6.222 |
| $\beta_{1,3}^s$ | 22 | 16 | 10 | 48 | 0.154 | 0.372 | 5/29 | -6.223 |
| $\beta_{1,4}^s$ | 24 | 18 | 12 | 54 | 0.113 | 0.333 | 11/65 | -6.228 |
| $\beta_{2,3}^s$ | 24 | 18 | 12 | 54 | 0.035 | 0.253 | 11/65 | -6.228 |
| $\beta_{2,4}^s$ | 26 | 20 | 14 | 60 | 0.011 | 0.238 | 1/6 | -6.232 |
| $\beta_{1,1,0}^s$ | 25 | 16 | 7 | 48 | 0.440 | 0.677 | 11/59 | -6.199 |
| $\beta_{1,1,2}^s$ | 29 | 20 | 11 | 60 | 0.218 | 0.442 | 13/73 | -6.214 |
| $\beta_{1,1,3}^s$ | 31 | 22 | 13 | 66 | 0.175 | 0.395 | 14/80 | -6.219 |
| $\beta_{1,2,3}^s$ | 33 | 24 | 15 | 72 | 0.094 | 0.316 | 15/87 | -6.223 |
| $\beta_{2,2,3}^s$ | 35 | 26 | 17 | 78 | 0.037 | | 16/94 | -6.227 |
| $\beta_{a1}^s$ | 9 | 6 | 1 | 16 | 0.556 | 0.871 | 1/5 | -6.196 |
| $\beta_{a2}^s$ | 11 | 8 | 3 | 22 | 0.313 | 0.614 | 5/27 | -6.219 |
| $\beta_{a3}^s$ | 11 | 8 | 3 | 22 | 0.287 | 0.569 | 5/27 | -6.210 |
| $\beta_{12}$[1] | 2 | 2 | 1 | 5 | | | 1/6 | -6.230 |
| $\chi_3$[1] | 2 | 2 | 0 | 4 | | | 1/5 | -6.245 |

**REFERENCE**

[1] B. J. Feng *et al.*, Nat. Chem. **8**, 564 (2016).